\documentclass[11pt,a4paper]{article}

\usepackage[margin=1in]{geometry} 
\usepackage{authblk}              
\usepackage{graphicx}             
\usepackage{booktabs}             
\usepackage{caption}              
\usepackage{hyperref}             
\usepackage{url}

\hypersetup{
    colorlinks=true,
    linkcolor=blue,
    filecolor=magenta,      
    urlcolor=blue,
    citecolor=blue,
}

\begin{document}

\title{\textbf{Structural Diversity Drives Disruptive Scientific Innovation}}

\author[1,2]{Yichun Peng\thanks{These authors contributed equally to this work.}}
\author[1,2]{Saike He$^\ast$}
\author[1]{Peijie Zhang$^\ast$}
\author[3]{Kang Zhao}
\author[1]{Yi Yang}
\author[1,2]{Ning Zhang}
\author[4]{Qingpeng Zhang}
\author[1,2]{Daniel Dajun Zeng\thanks{Corresponding authors. E-mail: zengdaniel@outlook.com (D.D.Z.); haopeng@cityu.edu.hk (H.P.)}}
\author[5]{Hao Peng$^\dagger$}

\affil[1]{\small State Key Laboratory of Multimodal Artificial Intelligence Systems, Institute of Automation, Chinese Academy of Sciences, Beijing 100190, China}
\affil[2]{\small University of Chinese Academy of Sciences, Beijing 101408, China}
\affil[3]{\small Department of Business Analytics, Tippie College of Business, The University of Iowa, Iowa City, IA 52242, United States of America}
\affil[4]{\small The University of Hong Kong, Institute of Data Science \& Department of Pharmacology and Pharmacy}
\affil[5]{\small Department of Data Science, City University of Hong Kong, Hong Kong SAR, China}

\date{\vspace{-5ex}} 

\maketitle

\begin{abstract}
Scientific innovation increasingly depends on collaboration, yet the organizational structure that fosters breakthrough ideas remains poorly understood. Existing metrics—such as team size or compositional diversity—capture readily observable characteristics but not the deeper architecture of collaboration. We introduce Structural Diversity (SD): the extent to which a team bridges multiple distinct knowledge communities within its prior collaboration network. Using a century-scale dataset of 260 million scientific publications (1900--2025) and combining causal inference with a quasi-natural experiment based on a U.S. National Science Foundation policy change in 2012, we show that SD is a powerful and robust predictor of disruptive innovation, outperforming traditional team novelty indicators such as team freshness and edge density. Moreover, SD positively interacts with team size and is able to mitigate the well-known ``curse of scale'' by transforming scale from a liability into a resource for creative synthesis. We find that one mechanism underlying this effect is Disciplinary Integration (DI): teams with higher SD can more effectively combine heterogeneous knowledge into novel configurations. Our findings position SD as both a new theoretical construct and an actionable design principle for organizing scientific collaboration. By linking the architecture of team assembly to the dynamics of creative discovery, our work offers a structural explanation for how collective intelligence can be systematically engineered to foster disruptive innovation.
\end{abstract}

\vspace{1em}

\vspace{2em}

\section*{Significance Statement}
Scientific innovation increasingly relies on team collaboration, yet the organizational architecture that sparks disruptive breakthroughs remain obscure. Analyzing 260 million scientific publications, we identify ``Structural Diversity'' (SD)—the extent to which a team bridges distinct knowledge communities—as a critical driver of disruptive innovation. Unlike traditional metrics that focus on team scale or member composition, SD captures the deep topology of creative synthesis. Crucially, high SD enables large teams to combine heterogeneous knowledge, mitigating the ``curse of scale'' where large teams typically become incremental. Our findings offer actionable design principles for organizing scientific collaboration. One policy implication is that fostering creative innovation depends more on deliberately designing teams that bridge structural holes than on simply assembling domain experts.

\newpage 

\section*{Introduction}

Scientific innovation is the engine of technological and societal progress, yet the organizational characteristics that nurture disruptive ideas remain elusive~\cite{bloom2020ideas}. Modern science is increasingly collaborative, making teams a widely used instrument for studying the mechanism of innovation performance~\cite{guimera2005team,wuchty2007increasing}. Existing research has addressed this question primarily through macro-level attributes such as team size~\cite{wu2019large}, cumulative experience~\cite{guimera2005team,jones2009burden}, disciplinary breadth~\cite{uzzi2013atypical}, or productivity~\cite{duch2012possible}. These aggregated indicators correlate with performance but can mask the intricate internal wiring of collaboration. They count the quantity of members within a team, not the quality of how team members are connected. As a result, we still have limited understanding of how teams access, combine, and reconfigure diverse perspectives — the process at the heart of scientific innovation.

Recent work has shifted focus inward to micro-level team features based on individual characteristics, such as examining diversity in demographics~\cite{alshebli2018preeminence}, workload distribution~\cite{haeussler2020division}, institutional background~\cite{jones2008multi}, or disciplinary mix~\cite{leahey2017prominent}. Yet these measures neglect the structural aspect of scientific collaboration in terms of who connects to whom and how different members are organized as a whole to effectively navigate the complex knowledge landscape. As a result, these metrics fall short in adequately capturing the diverse sources of perspectives among team members. For instance, ``team freshness''~\cite{zeng2021fresh}—the proportion of previously unconnected collaborators—captures novelty in team assembly but ignores the previous experience of incumbents and the broader context of network connectivity. A team of strangers drawn from the same intellectual lineage can be cognitively homogeneous, whereas a same-institution team that spans distinct knowledge communities can bring cognitive variety that may lead to more creative solutions. 

Similarly, traditional diversity measures based on gender or ethnicity would treat teams composed of members from a single demographic group or social identity as exhibiting little variation, despite potentially meaningful differences along other cognitive dimensions.

\begin{figure}[htbp] 
    \centering
    \includegraphics[width=1\linewidth]{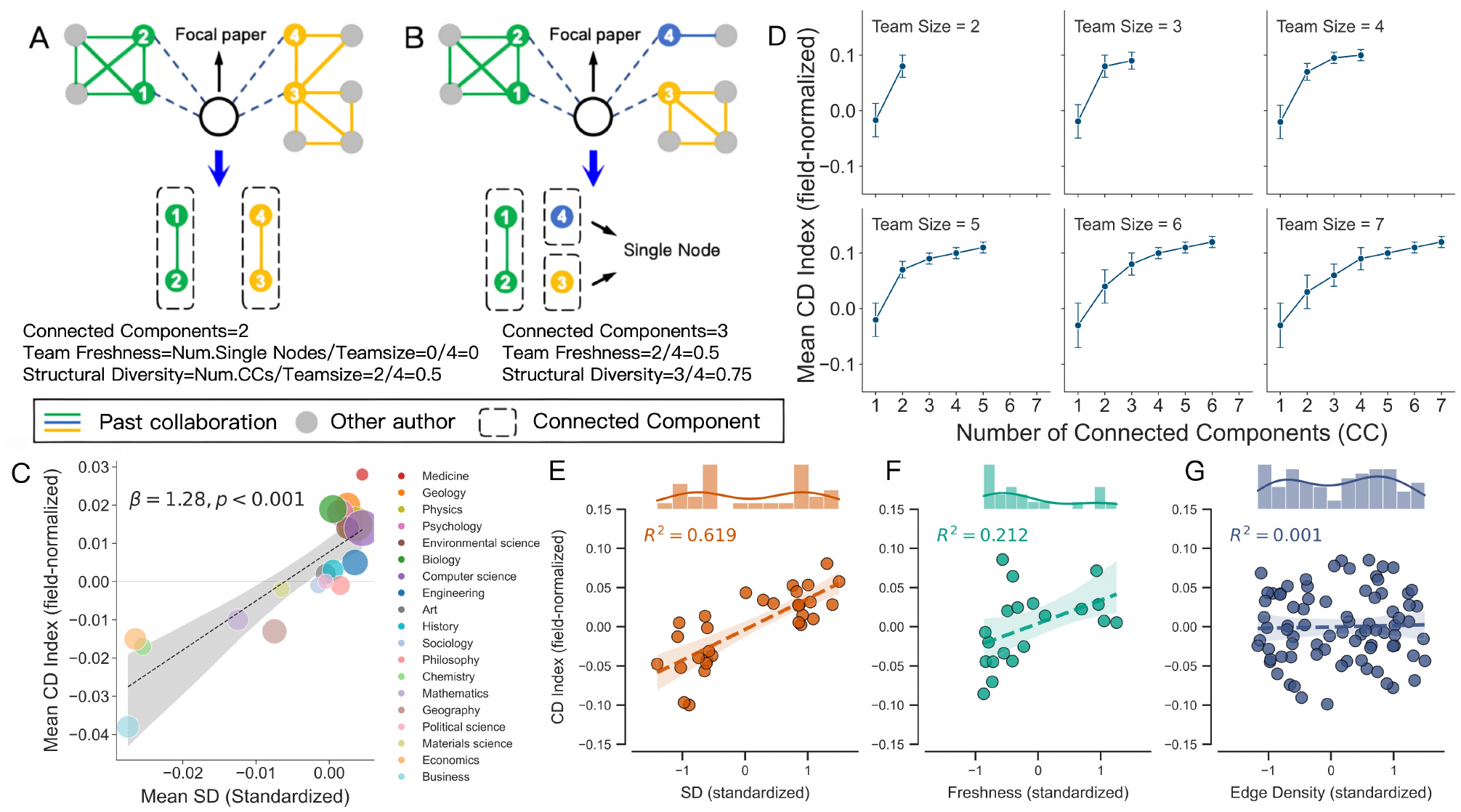}
    \caption{\textbf{SD as a robust predictor of disruptive scientific innovation.} (A-B) Illustration of how SD captures the potential of distinct knowledge communities for innovation. (C) Each point represents one discipline (19 in total). The dashed line shows a statistically positive correlation between the mean SD (standardized) and the mean CD Index (field-normalized) across all areas of science. Bubble size is proportional to the number of papers in each discipline. (D) Relationship between the number of CCs within a team's historical co-authorship network and the CD Index, stratified by a paper's team size from 2 to 7 authors (N = $256,600,000$ papers). Each subplot shows that, holding team size constant, more CCs are associated with higher disruptive impact. Error bars represent $95\%$ confidence intervals. (E-G) Mean CD Index as a function of SD, team freshness, and edge density. Points show the mean values for equal-sized bins of papers. Each colored regression line represents the simple linear fit of these binned averages ($R^2$ refers to the goodness of fit for the binned data). Histogram and kernel density curve at the top of each subplot show the distribution of the respective (standardized) metric.}
    \label{fig:Fig1}
\end{figure}

To address these limitations, we take a meso-level perspective centered on structural diversity (SD) — the extent to which a team weaves together multiple distinct knowledge communities. Inspired by prior literature on the power of structural holes in social networks~\cite{ugander2012structural, burt2004structural, ahuja2000collaboration}, we operationalized SD as the weighted number of connected components (CCs) based on the past collaboration network of team members (see details in Methods). 

Our SD index quantifies the degree of cognitive heterogeneity of team members in the knowledge space of scientific research. Our measure goes beyond simple group size or team freshness that merely counts the scale of team or the ratio of new members without considering potential intellectual and field differences among group members (Fig.~\ref{fig:Fig1}\textit{A-B} shows an illustration of their differences).

To evaluate the predictive power of our SD measure, we analyze a century-scale corpus of 260 million scientific papers (1900--2025) across 19 disciplines indexed in the OpenAlex database~\cite{priem2022openalex} (see details in Methods). 
We first demonstrate that SD is a robust predictor of disruptive innovation at both the discipline level and the paper level. We find that our SD measure significantly outperforms traditional team novelty indicators such as team freshness and edge density, allowing us to make better predictions about when and why scientific collaboration becomes most transformative.
We show that the predictive power of SD is robust even after taking into account key confounding factors including linguistic content, author experience, member novelty, and team size. This suggests that even for teams with a fixed size, one can still maximize their innovation potential through high structural diversity. 

We then delineate boundary conditions by examining the variational effects of SD on disruption across disciplines and its interaction effect with team size. Surprisingly, we find a statistically positive interaction effect between structural diversity and team size, indicating that the innovation benefit of SD is even higher for larger teams. Our finding suggests that SD can be effectively leveraged to mitigate the well-known ``curse of scale''~\cite{wu2019large} in scientific innovation by turning team scale from a liability into a resource for creative synthesis.

To further support these empirical findings, we employ a dual causal inference strategy combining the Propensity Score Matching (PSM)~\cite{rosenbaum1983central} and a quasi-natural experiment based on a U.S. National Science Foundation (NSF) policy change implemented in 2012. 
Finally, we explore one underlying behavioral mechanism, revealing that SD drives disruptive innovation by fostering disciplinary integration—the effective recombination of heterogeneous knowledge domains in new ways. Together, our results establish SD as both a theoretical construct and a practical design principle for facilitating disruptive innovation through the architecture of collective intelligence.

\section*{Results}

\subsection*{SD Predicts Disruptive Innovation Above and Beyond Team Scale and Member Novelty}

\begin{table}[htbp]
\centering
\caption{OLS regression of Structural Diversity (SD) on the field-normalized CD Index.}
\label{tab:regression_models}
\begin{tabular}{l c c c}
\toprule
\multicolumn{4}{c}{\textit{Dependent Variable: CD Index (field-normalized)}}  \\
\addlinespace[0.5ex]
& Model 1 & Model 2 & Model 3 \\
\midrule
SD (standardized) & $0.026$ ($p < 0.001$) & $0.026$ ($p < 0.001$) & $0.025$ ($p < 0.001$) \\
\addlinespace
\textbf{Content Controls} & & & \\
\hspace{1em}Title Word Count & & $-0.002$ ($p < 0.001$) & $-0.002$ ($p < 0.001$) \\
\hspace{1em}Title Readability Score & & $0.001$ ($p < 0.055$) & $0.001$ ($p = 0.070$) \\
\hspace{1em}Title Promotional Words (\%) & & $0.016$ ($p < 0.001$) & $0.012$ ($p < 0.001$) \\
\addlinespace
\textbf{Team/Author Controls} & & & \\
\hspace{1em}Team Size (log) & & & $-0.013$ ($p < 0.001$) \\
\hspace{1em}Team Freshness & & & $0.004$ ($p < 0.001$) \\
\hspace{1em}Author Career Age & & & $0.002$ ($p = 0.059$) \\
\hspace{1em}Author Career Age$^{2}$ & & & $-0.050$ ($p = 0.006$) \\
\hspace{1em}Author Institution h-index & & & $0.001$ ($p = 0.066$) \\
\hspace{1em}Author No. Publications (log) & & & $-0.004$ ($p < 0.001$) \\
\addlinespace
\textbf{Fixed Effects Controls} & & & \\
\hspace{1em}Discipline (19 categories) & Yes & Yes & Yes \\
\hspace{1em}Publication Year (1900--2025) & Yes & Yes & Yes \\
\midrule
\textbf{Model Statistics} & & & \\
Observations & $260,400,000$ & $260,400,000$ & $260,400,000$ \\
$R$-squared & $0.009$ & $0.010$ & $0.014$ \\
\bottomrule
\end{tabular}
\medskip
\parbox{\linewidth}{\footnotesize \textbf{Note:} Variables such as Team Size and Author No. Publications are log-transformed to account for their skewed distributions and diminishing marginal returns. We include a squared term for Author Career Age to capture its non-linear effect on innovation behaviors such as authors' self-promotion~\cite{peng2025gender} and depth of knowledge search~\cite{cui2025agingnarrowingscientificinnovation}. Variable measurements are shown in Methods.}
\end{table}

We define SD as the ratio of connected components (CCs) to the total number of team members based on the team's historical co-authorship network prior to a focal publication (Fig.~\ref{fig:Fig1}\textit{A} and \textit{B}; see details in Methods). SD is designed to capture the extent to which the team bridges distinct, pre-existing co-authorship communities, serving as a proxy for latent cognitive and institutional heterogeneity. 
To examine whether our SD measure holds any predictive power, we analyze a century-scale dataset of over 260 million research papers and quantify their degree of disruptiveness using the field-normalized Consolidation–Disruption (CD) Index~\cite{wu2019large,funk2017dynamic} (see details in Methods). To ensure consistency and facilitate comparison with existing measures, we employ z-score standardization for SD in all subsequent analyses.

First, at the macro level when treating each discipline as a unit of analysis, we find a statistically high correlation between the mean SD of teams within a discipline and its average CD Index ($\beta = 1.28$, $p < 0.001$; Fig.~\ref{fig:Fig1}\textit{C}). In addition, Fig.~\ref{fig:Fig1}\textit{C} shows that disciplines characterized by a higher average SD also exhibit more innovation output (Spearman rank correlation between mean SD and publication volume: $\rho = 0.355$, $p < 0.001$), suggesting a systemic link between collaboration patterns and a field's macroscopic innovative capacity.

This finding is similarly corroborated at the paper level. Fig.~\ref{fig:Fig1}\textit{D} shows that, for teams of fixed size, a higher number of connected components is consistently linked with a higher degree of disruptive innovation. Furthermore, in predicting the field-normalized CD Index, SD consistently outperforms traditional team novelty metrics such as team freshness~\cite{zeng2021fresh} and edge density~\cite{ahuja2000collaboration}. The advantage of SD is visually evident in Fig.~\ref{fig:Fig1}\textit{E-G}: the simple binned regression model based on SD explains substantially more variance in CD Index ($R^2= 0.619$) than that based on team freshness ($R^2= 0.212$) or edge density ($R^2 = 0.001$).

Table~\ref{tab:regression_models} shows that the positive effect of SD on disruptive innovation remains strong even after controlling for confounding factors related to paper's linguistic content, author's research experience, and team-level characteristics, as well as fixed-effects for publication year and research fields ($\beta = 0.025$, $p < 0.001$; see variable measurement in Methods). The statistical association between SD and CD Index is also robust in three different model specifications (Table~\ref{tab:regression_models}).

Furthermore, the main effect of SD on disruptive innovation remains positive and statistically significant across a series of robustness checks. Specifically, results hold under alternative operationalizations of SD (unstandardized SD or CC-per-member ratio; see details in \textit{SI Appendix}, Table S1), alternative innovation measures (Novelty Index; \textit{SI Appendix}, Table S1), author disambiguation (\textit{SI Appendix}, Table S2), the exclusion of very large teams ($>$15 members, \textit{SI Appendix}, Table S2), and varying pre-history windows for computing CC and SD (3, 4, 6, and 7 years, \textit{SI Appendix}, Table S3). These consistent findings confirm that the positive statistical relationship between SD and CD is not sensitive to different measurement definitions, model specifications, or sample selections. 

The strength of SD lies in something that goes beyond team novelty metrics and network cohesion descriptors. We find that, conditional on SD, adding edge density and clustering coefficient into the regression model shown in Table~\ref{tab:regression_models} yields only marginal incremental gains (1.4\% relative increase in $R^2$; \textit{SI Appendix}, Table S4). However, when incorporating SD into a regression model that includes edge density and clustering coefficient, and the same set of controls used in Table~\ref{tab:regression_models}, SD significantly enhances the model's explanatory power, increasing $R^2$ by $16.7\%$ (\textit{SI Appendix}, Table S5).

Taken together, our finding of SD's predictive power, robustness checks, and comparison of model performance with alternative measures collectively establish SD as a robust meso-level predictor of disruptive innovation, whose strength goes beyond traditional group features including team size and membership novelty.

\subsection*{Boundary Conditions and the Mitigation of Team Scale}

\begin{figure}[htbp] 
    \centering
    \includegraphics[width=1\linewidth]{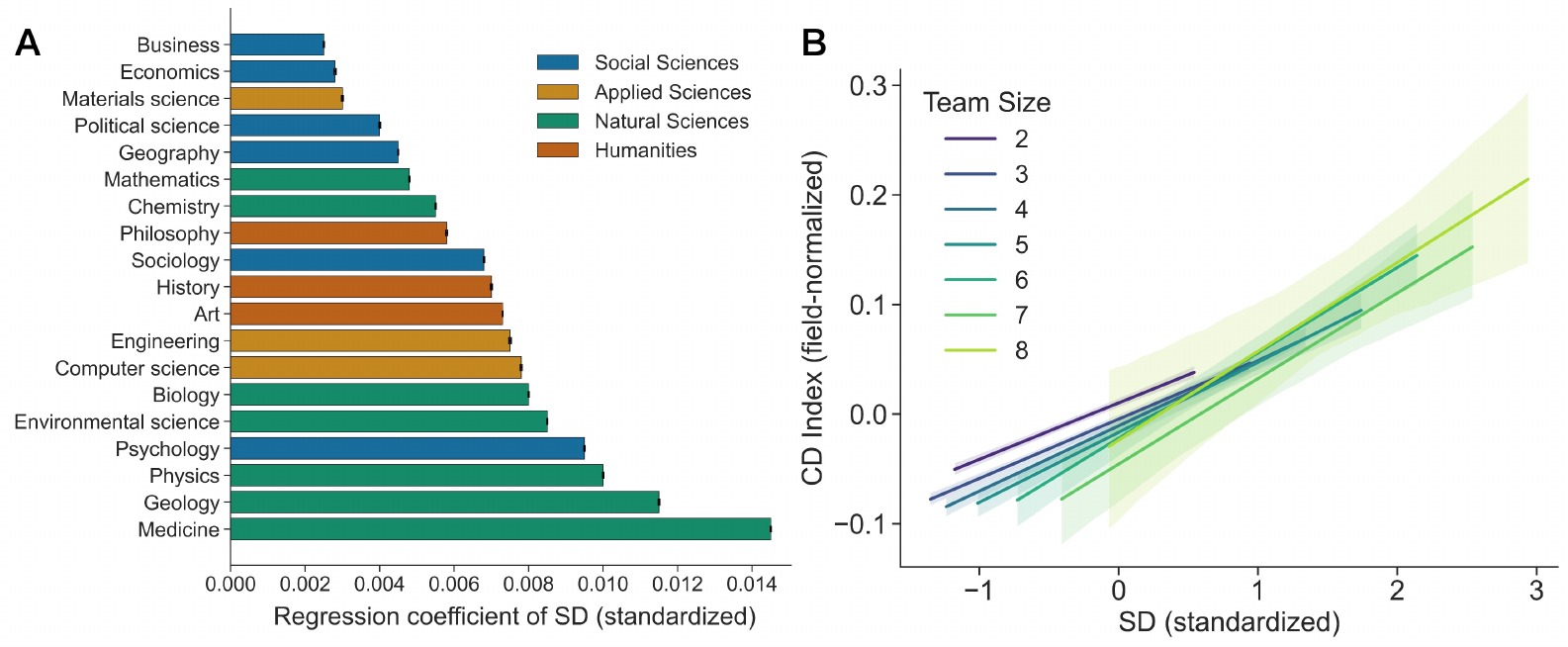}
    \caption{\textbf{Disciplinary heterogeneity of SD and the mitigation of ``Curse of Scale''.} (A) Heterogeneous effects of SD on disruptive innovation across 19 disciplines. Each horizontal bar shows the regression coefficient ($p < 0.001$ for all 19 disciplines) between SD (standardized) and CD Index (field-normalized) by fitting the same regression model in Table~\ref{tab:regression_models} to all papers in each discipline. The effect of SD is strongest in natural sciences (e.g., Medicine, Geology, and Physics) and weakest in social sciences such as Business and Economics. (B) Positive interaction effect between SD (standardized) and Team Size (see regression coefficients in Table~\ref{tab:interaction_effect}). The margins plot shows the predicted CD Index (field-normalized) as a function of SD (standardized) for different team sizes. The slope becomes steeper as team size increases (from purple to light green), indicating that SD can effectively mitigate the ``curse of scale'' by amplifying the disruptive potential of large teams. Error bars represent $95\%$ confidence intervals.}
    \label{fig:Fig2}
\end{figure}

\begin{table}[htbp]
\centering
\caption{OLS regression reveals a statistically positive interaction effect between Structural Diversity (SD) and Team Size on CD Index.}
\label{tab:interaction_effect}
\begin{tabular}{l c}
\toprule
\multicolumn{2}{c}{\textit{Dependent variable: CD Index (field-normalized)}} \\
\midrule
\textbf{Main Predictors} & \\
SD (Standardized) & $0.016$ ($p < 0.001$) \\
Team Size (log) & $-0.031$ ($p < 0.001$) \\
\addlinespace
\textbf{Interaction} & \\
SD $\times$ Team Size (log) & $0.017$ ($p < 0.001$) \\
\addlinespace
\textbf{Content Controls} & \\
Title Word Count & $-0.002$ ($p < 0.001$) \\
Title Readability & $0.001$ ($p = 0.084$) \\
Title Promotional Words (\%) & $0.009$ ($p < 0.001$) \\
\addlinespace
\textbf{Team/Author Controls} & \\
Team Freshness & $0.002$ ($p < 0.001$) \\
Author Career Age & $0.002$ ($p = 0.049$) \\
Author Career Age$^{2}$ & $-0.098$ ($p = 0.013$) \\
Author Institution h-index & $0.003$ ($p = 0.015$) \\
Author No. Publications (log) & $-0.006$ ($p < 0.001$) \\
\addlinespace
\textbf{Fixed Effects Controls} & \\
Discipline (19 categories) & Yes \\
Publication Year (1900--2025) & Yes \\
\midrule
\textbf{Model Statistics} & \\
Observations & $260,400,000$ \\
$R$-squared & $0.014$ \\
\bottomrule
\end{tabular}
\end{table}

Our analysis in the previous section identifies Structural Diversity (SD) as a consistent and robust predictor of scientific disruption, possibly operating through the integrative combination of distinct knowledge communities. We next examine whether the power of SD is universal across different research fields and how its effect may vary across common features of team assembly such as the group size.

First, we find that the magnitude of correlation between SD and CD Index varies meaningfully across disciplines. Fig.~\ref{fig:Fig2}\textit{A} shows that, contrary to the common assumption that social sciences and humanities are inherently integrative~\cite{cole1983hierarchy}, the strongest positive association appears in fields of Natural Sciences, including Medicine ($\beta = 0.014$, $p < 0.001$), Geology ($\beta = 0.012$, $p < 0.001$), and Physics ($\beta = 0.010$, $p < 0.001$), where epistemic boundaries are more codified and rigid than other fields~\cite{peng2021neural}. Conversely, the marginal benefit of additional SD diminishes in conceptually more soft and pluralistic areas such as Social Sciences (mean $\beta = 0.004$, $p < 0.001$), Applied Sciences (mean $\beta = 0.006$, $p < 0.001$), and Humanities (mean $\beta = 0.007$, $p < 0.001$).

Second, we examine if the predictive power of SD on disruptive innovation is uniform across different team sizes by adding an interaction term between SD and team size in our previous OLS regression. We find a statistically positive interaction effect between SD and team size ($\beta = 0.017$, $p < 0.001$; Table~\ref{tab:interaction_effect}), suggesting that the effect size of SD on innovation performance is even greater for larger teams. 
Fig.~\ref{fig:Fig2}\textit{B} further illustrates how SD interacts with team scale. Specifically, large teams generally suffer from the ``curse of scale,'' where increased manpower correlates with reduced disruptiveness~\cite{wu2019large}. Table~\ref{tab:interaction_effect} indeed shows that each additional team member is associated with reduced CD Index ($\beta = -0.031$, $p < 0.001$). 

However, the positive correlation between SD and disruptive innovation becomes even more pronounced as the team size increases, as the slope of the marginal prediction becomes steeper. 
In fact, Fig.~\ref{fig:Fig2}\textit{B} shows that, holding controls constant, larger teams can become more disruptive than smaller ones when their SD is optimized, suggesting that SD can be effectively leveraged to mitigate the diminishing returns typically associated with large teams. 
Our finding thus reveals that the team scale itself is not inherently conservative; rather, structural homogenization may be the curse limiting team creativity.

\subsection*{Robustness Test Using Propensity Score Matching}

\begin{figure}[htbp] 
    \centering
    \includegraphics[width=1\linewidth]{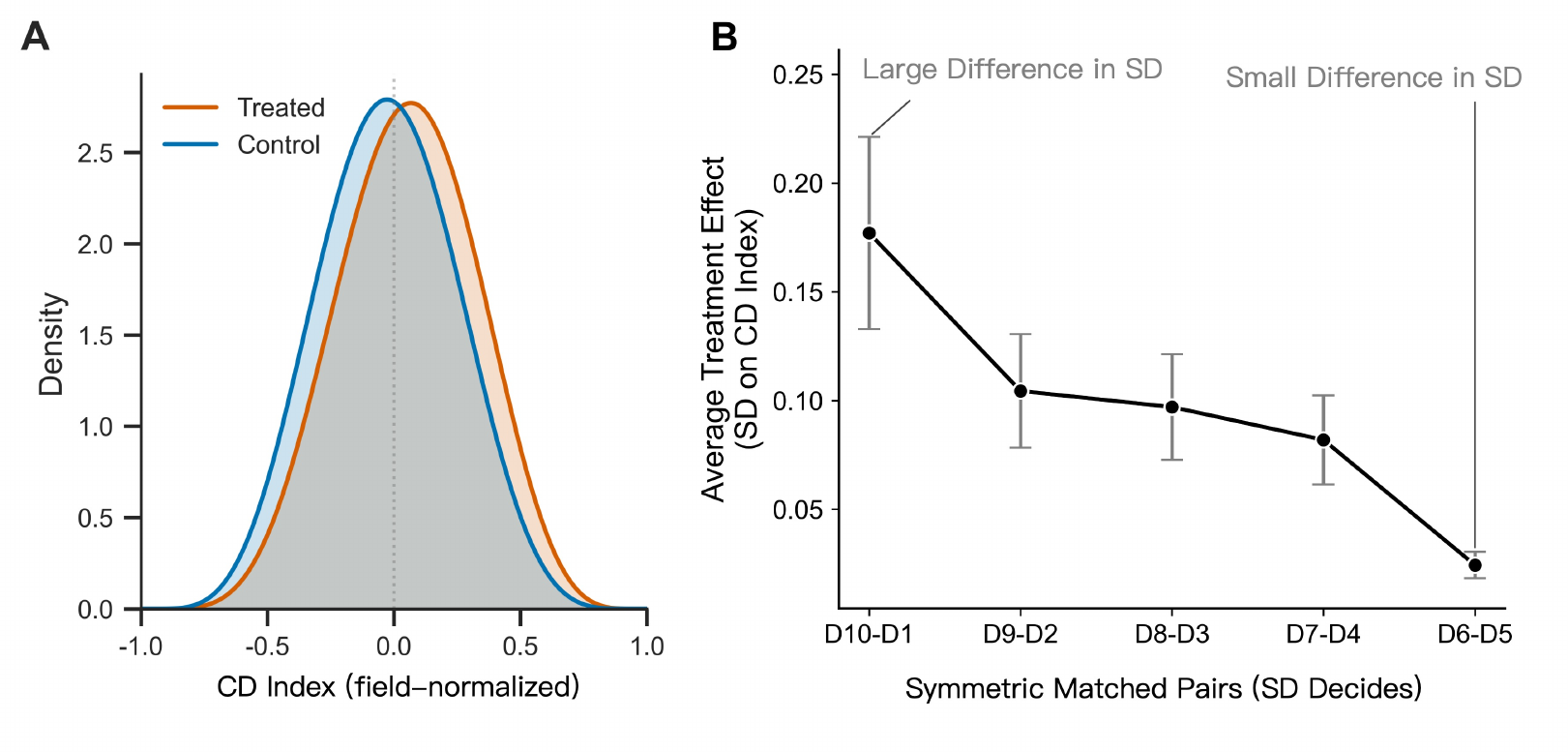}
    \caption{\textbf{Robustness test using the Propensity Score Matching framework further supports the positive association between SD and disruptive innovation.} The analysis is based on $690,972$ Computer Science papers. (A) Kernal density estimation of the CD Index for treated teams (Top quartile SD, orange) and their matched control teams (Bottom quartile SD, blue). The distribution of CD Index for high-SD teams is statistically right-shifted than that for low-SD teams (mean CD: $0.06$ vs. $-0.06$, $p<0.001$), indicating a large increase in team disruptiveness. (B) Robustness checks using a symmetric interval matching approach based on ten deciles of SD. The ``D1'' pair represents the comparison between the most structurally diverse teams (Top $10\%$) and the least diverse teams (Bottom $10\%$), whereas the ``D5'' pair compares two sets of matched teams with SD near the median. The dots represent point estimates of the Average Treatment Effect on the Treated (ATT) of SD on CD Index. The positive effect of SD is substantial at the two extremes but converges to zero for teams with the average level of structural diversity. Error bars indicate $95\%$ bootstrapped confidence intervals.}
    \label{fig:Fig3}
\end{figure}

Our next objective is to assess whether our main results are sensitive to differences in observed covariates between different papers using the standard causal inference framework.
Given the immense computational and memory demands of causal inference, analyzing the complete OpenAlex dataset (about 260 million papers) is computationally prohibitive. We therefore narrow our analytical scope to the AMiner dataset consisting of $690,972$ papers in Computer Science~\cite{tang2008arnetminer} (see Methods). As shown in Fig.~\ref{fig:Fig1}\textit{C} and Fig.~\ref{fig:Fig2}\textit{A}, Computer Science not only provides sufficient research outputs but also exhibits a statistically positive effect of SD on disruptive innovation, making it a suitable discipline for testing the causal effect of SD.

We use Propensity Score Matching (PSM)~\cite{rosenbaum1983central} as a robustness check to examine whether and to what extent a team's SD is statistically associated with the disruptiveness of its scientific innovation (CD Index). 
The propensity score of a team's SD is estimated based on all content, author, and team related variables and fixed effects for year (listed in Table~\ref{tab:regression_models}). We convert the continuous variable SD (standardized) into four quartiles and use teams whose SD ranks in the top quartile of papers as the treatment group. We use the one-to-one matching technique to identify control units with the most similar covariate values from the symmetric bottom quartile of papers. The matching is performed under the Stable Unit Treatment Value Assumption (SUTVA)~\cite{rubin1980randomization}. Balance diagnostics, including the standardized mean difference ($SMD < 0.1$ for all covariates), confirm that a high degree of balance has been achieved between the matched groups.

Fig.~\ref{fig:Fig3}\textit{A} shows that the high-SD ``treated'' teams exhibit a statistically higher tendency towards disruptive innovation compared to the low-SD ``control'' teams (mean field-normalized CD Index: $0.06$ vs. $-0.06$, $p<0.001$). This indicates that, after restricting the analysis to papers with comparable observed characteristics, higher SD significantly increases the degree of a team making disruptive scientific contributions. 

To ensure the robustness of this result, we split all teams or papers into more granular groups based on ten deciles and apply the same PSM framework by gradually changing the symmetric matching interval, ranging from the most different pairs (d1: Top $10\%$ vs. Bottom $10\%$) to those near the median (d5: Top $40\%-50\%$ vs. Bottom $40\%-50\%$). Fig.~\ref{fig:Fig3}\textit{B} indicates that the estimated Average Treatment Effect on the Treated (ATT) of SD on CD consistently decreases as the degree of structural difference between the two matched groups narrows. This suggests that teams bridging the most distant knowledge communities (i.e., those with the highest SD) can produce the most innovation gains.

\subsection*{A Quasi-Natural Experiment Reinforces the Positive Role of SD}

To provide a quasi-experimental stress test on the association between SD and disruptive innovation, we leverage a policy change introduced by the U.S. National Science Foundation (NSF). The policy, which expanded funding eligibility for multi-institutional projects in 2012, generated an exogenous condition that encouraged broader collaboration across U.S. research institutions. 

We again exclusively use the AMiner dataset, as it provides the specific NSF funding metadata required to identify projects affected by the policy change, along with high-quality author disambiguation data needed for accurately constructing the collaboration network. Using the AMiner dataset filtered for NSF-funded papers ($N = 41,458$), we adopt a pre–post design, comparing team structure and innovation outputs in the two years before (2010–2011, $N=18,267$) and the two years after (2012–2013, $N=23,191$) the policy implementation.

\begin{figure}[htbp] 
    \centering
    \includegraphics[width=1\linewidth]{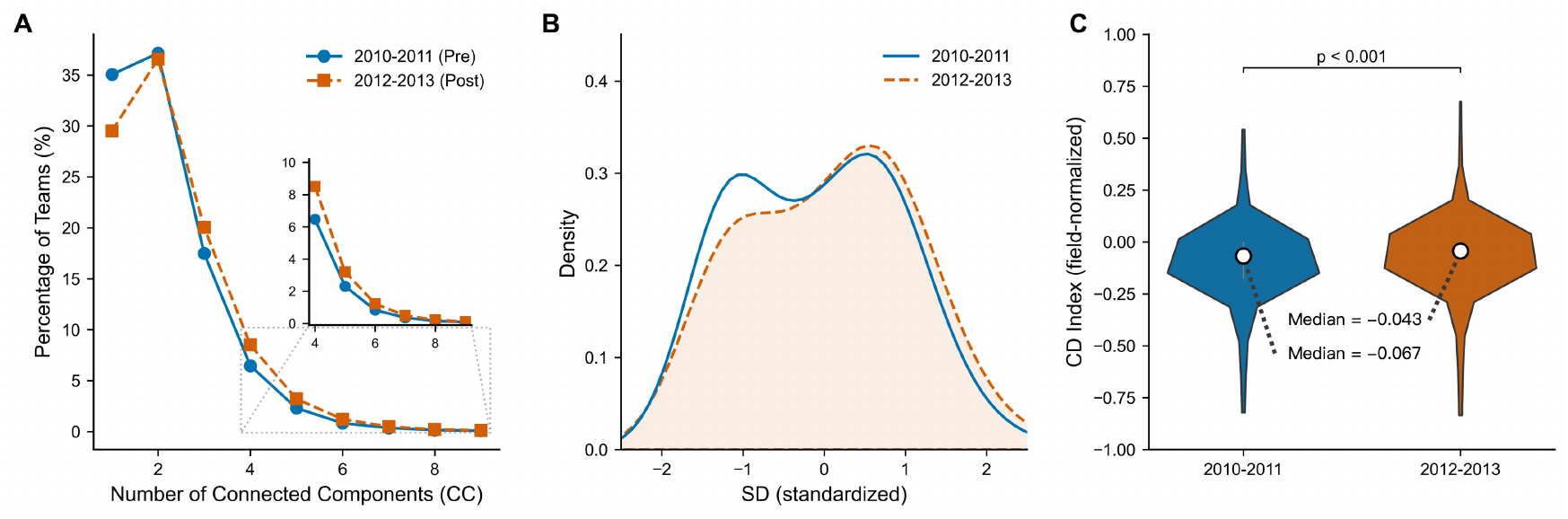}
    \caption{\textbf{A quasi-natural experiment based on the NSF policy change in 2012 supports the correlation between SD and disruptive innovation.} (A) The percentage of teams by the number of Connected Components (CC) before (2010-2011, $N = 18,267$) and after (2012-2013, $N = 23,191$) the NSF policy implementation. All papers ($N = 41,458$) were NSF-funded in the AMiner dataset. The inset highlights a sharp increase in teams with high CC counts (right tail) in the post-policy period (Mann-Whitney U test, $p < 0.001$). (B) Kernel density estimation of SD shows a distributional shift towards higher structural diversity after the policy change (Mann-Whitney U test, $p < 0.001$). (C) Violin plots comparing the CD Index before and after the policy implementation. The median (or mean) CD Index increased statistically in the post-policy period ($p < 0.001$ in a two-sided t-test), suggesting that the policy-induced structural diversification translated into more creative research outputs.}
    \label{fig:Fig4}
\end{figure}

Fig.~\ref{fig:Fig4}\textit{A} shows that the policy triggered a noticeable increase in team-level structural diversity. The average number of connected components increases from $2.10$ to $2.28$ ($p < 0.001$ based on t-test or Mann–Whitney U test), with the largest proportional changes occurring in the right tail of the distribution (Inset in Fig.~\ref{fig:Fig4}\textit{A}). Fig.~\ref{fig:Fig4}\textit{B} shows that the distribution of team SD also experienced an increasing trend after the policy change ($p < 0.001$ based on Mann–Whitney U test). 
These patterns indicate that this NSF policy effectively catalyzed more boundary-spanning scientific collaborations. 

We then assess whether this policy-induced increase in SD translated into greater innovation outcomes. 
Fig.~\ref{fig:Fig4}\textit{C} shows that the policy's boost on innovation is evident: both the median and the mean of CD Index increase statistically in the post-policy period ($p < 0.001$ based on the independent sample t-test). This result suggests that this NSF policy led to more disruptive innovation likely because it encouraged more structurally diverse teams in scientific collaboration.
We do not interpret this natural experiment as providing a definitive causal estimate. Rather, it serves as a robustness check reassuring that the empirically observed association between SD and disruption persists under a plausibly exogenous condition.

\subsection*{SD Fosters Disruptive Innovation Through Disciplinary Integration}

\begin{figure}[htbp] 
    \centering
    \includegraphics[width=0.9\linewidth]{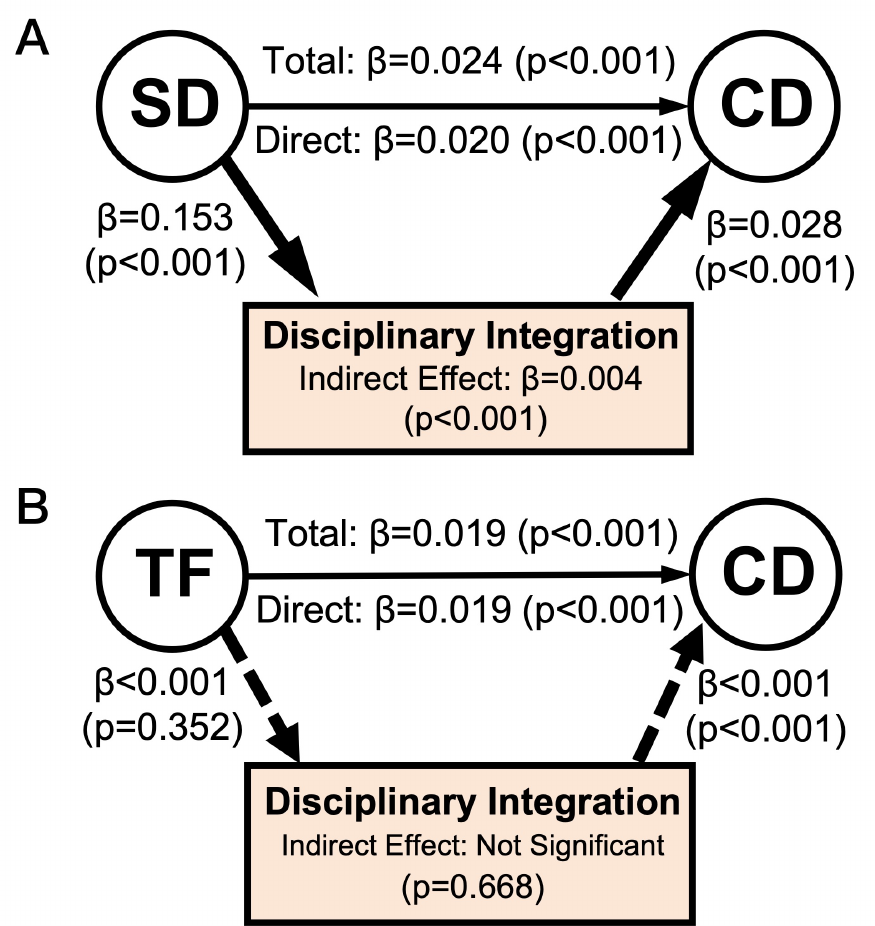}
    \caption{\textbf{Disciplinary Integration (DI) mediates a sizable fraction of the effect of SD on disruptive innovation.} (A) Path diagram linking team SD to their CD Index via disciplinary integration. SD significantly predicts DI ($\beta = 0.153$, $p < 0.001$), which in turn predicts CD Index ($\beta = 0.028$, $p < 0.001$). About $18\%$ of the effect of SD on CD (total effect size: $0.024$, $p < 0.001$) can be explained by DI (indirect effect size: $0.004$, $p < 0.001$). (B) The same path diagram for Team Freshness (TF). Unlike SD, TF shows no predictive power on DI. The indirect path to innovation via DI is also insignificant, demonstrating that the unique power of SD lies in team members' capacity to synthesize distinct knowledge domains. The mediation analysis is conducted based on N = $260,400,000$ papers indexed in the OpenAlex database. All regression models in this analysis further control for all confounding variables except Team Freshness listed in Table~\ref{tab:regression_models}.}
    \label{fig:Fig5}
\end{figure}

The core value of SD lies in its ability to position teams as knowledge brokers~\cite{burt2004structural} that connect previously unlinked intellectual communities. This structural bridging necessarily requires the recombination of heterogeneous ideas. We therefore introduce and hypothesize Disciplinary Integration (DI)~\cite{porter2009science} as a key mediating construct that captures the behavioral manifestation of knowledge recombination. Unlike the counting of superficial multidisciplinary labels, DI reflects the actual diversity of knowledge sources utilized by the team. It is defined as the extent to which a paper effectively synthesizes references from disparate disciplinary domains (see details in Methods). Conceptually, we propose the following mediation pathway: SD (structure) → DI (behavior) → CD (innovation outcome) as one of the potential mechanisms behind the power of SD.

Leveraging the comprehensive OpenAlex dataset (N = $260,400,000$) based on a formal mediation analysis, we find that teams with greater SD indeed exhibit significantly higher levels of DI ($\beta = 0.153$, $p < 0.001$; Fig.~\ref{fig:Fig5}\textit{A}). This result supports the theoretical premise that SD not only connects disparate communities but also translates these new connections into genuinely cross-disciplinary knowledge synthesis. Furthermore, Fig.~\ref{fig:Fig5}\textit{A} shows that DI itself strongly predicts innovation disruptiveness ($\beta = 0.028$, $p < 0.001$). Thus, Disciplinary Integration could function as a behavioral conduit linking team structure to creative performance in science production.

Fig.~\ref{fig:Fig5}\textit{A} shows that a sizable portion, about $18\%$, of SD's total effect on innovation is contributed indirectly through DI ($p < 0.001$; \textit{SI Appendix}, Table S6). This result suggests that DI is one of the primary mechanisms through which SD fosters disruptive innovation. 
Importantly, when we substitute SD with team freshness, its indirect effect via DI is not significant ($p = 0.668$; Fig.~\ref{fig:Fig5}\textit{B}).
This result underscores that knowledge integration is the distinctive channel through which SD drives creativity and explains the unique advantage of SD over mere member novelty.

\section*{Discussion}

\subsection*{Beyond ``Interdisciplinary Labels'': Structured Recombination of Tacit Knowledge}

For a long time, science policy has often advocated for and promoted interdisciplinary research through disciplinary labels (e.g., funding ``biophysics'' projects). However, our findings suggest that this label-based classification may be misleading~\cite{peng2021neural}. The SD metric reveals that the true driver of disruptive innovation comes not from the patchwork of disciplinary names, but from the structured recombination of tacit knowledge. A team composed of a physicist and a biologist, if they have long occupied the same social circle (low SD), often already share similar thinking paradigms, resulting in ``pseudo-interdisciplinarity.'' In contrast, high-SD teams break this echo chamber effect by introducing members from different ``epistemic lineages.'' This implies that the scientific community needs to redefine ``diversity'': it is not a mixture of identity labels, but a topological spanning of the academic social networks.

\subsection*{The Power of ``Weak Ties'' and the Evolutionary Game of Science}

From an evolutionary perspective, scientific development can be viewed as a dynamic interplay between ``Exploitation'' and ``Exploration.'' Existing research structures (such as laboratory heritage and mentorship) tend to build ``strong tie'' networks, which are conducive to the inheritance and deepening of knowledge (Exploitation). However, our SD model demonstrates the applicability of Granovetter's ``Strength of Weak Ties'' theory in scientific discovery: disruptive mutations are more often to occur at the edges and gaps of the network. 

We findings also enrich the current literature on team science. Large teams are usually considered bastions of conservatism, but our study proposes a counter-intuitive correction: as long as sufficient structural heterogeneity is introduced, large teams can transform from ``redundant bureaucracies'' into ``rich ecosystems.'' This indicates that future ``Big Science'' projects should not merely pursue expansion in scale, but should deliberately incorporate ``Structural Holes'' into the internal design, purposefully creating opportunities for bridging cognitive segregation and resolving differences within the team to prevent entropy loss caused by cognitive homogenization.

\subsection*{The Trade-off between Collaboration Efficiency and Disruption: Structural Diversity as ``Cognitive Friction''}

Our findings imply a fundamental trade-off in scientific organization: the tension between execution efficiency and disruptive potential. We speculate that the mechanism underlying this trade-off is Cognitive Friction. Bridging disconnected knowledge communities (high SD) inevitably entails higher communication costs and difficulties in consensus formation. However, this structural ``inefficiency'' and conflict-resolving process is precisely the hotbed of revolutionary innovation. Traditionally cohesive teams (low SD) are highly efficient at executing ``Normal Science'' tasks because they possess shared language and tacit understanding; whereas high-SD teams must establish new understandings first through conflict and negotiation within heterogeneous mental models (referred to as Disciplinary Integration, DI, in the text). Therefore, our results suggest that if the current research funding system excessively pursues ``short-term output efficiency'' and ``team harmony,'' it risks systematically pushing the ecosystem towards incrementalism and unintentionally eliminating those high-friction but high-return teams.

\subsection*{Implications for Artificial Intelligence and the Future of Science}

Finally, as the AI for Science paradigm becomes increasingly prevalent, our findings provide a unique ``human'' perspective. AI excels at processing massive data (scale advantage), but current AI models are mostly replicating statistical patterns embodied in existing corpora, essentially performing the most efficient ``Consolidation'' process. The unique value of human scientists lies in the unpredictability of their constantly evolving social networks—weaving them into disparate communities and fields through accidental encounters and structural bridging. Our SD concept thus to some degree quantifies the potential of this ``Engineered Serendipity.'' In the future of science shaped by human-machine symbiosis, machines will optimize efficiency and integration, while humans should focus on building and cultivating diverse social structures to venture into those knowledge wildernesses beyond the reach of algorithms.

\section*{Summary}

Our study introduces and empirically validates the new concept of structural diversity (SD) as a powerful and robust predictor of disruptive innovation, outperforming traditional metrics such as team freshness and other network cohesion metrics. By quantifying the degree to which teams bridge disconnected knowledge communities, we show that SD captures the structural precondition for transformative creativity in scientific collaboration. Mechanistically, SD operates through disciplinary integration (DI) — a behavioral pathway that compels teams to recombine heterogeneous knowledge components into novel configurations, thus producing more disruptive outcomes. Beyond its core effect, SD exhibits amplification in larger teams and systematic variations across disciplines, revealing when and where diversity yields the highest innovation return. Collectively, these results establish SD not merely as a descriptive property of social networks but as an actionable design principle for organizing scientific collaboration. By linking the architecture of social structure to the dynamics of discovery, our work provides both conceptual insight and practical guidance for building a more innovative and resilient knowledge ecosystem.

\section*{Methods}

\subsection*{Data Sources}
Our study uses the OpenAlex bibliographic database~\cite{priem2022openalex} (2025 version), consisting of 260,400,000 scientific papers published between 1900 and 2025. Each paper contains detailed metadata including paper title, author and affiliation information, publication year, venue, and research topics.
These papers are categorized into 19 disciplines based on the top-level fields defined in the OpenAlex database. To examine disciplinary heterogeneity of SD on innovation performance, we aggregate 19 disciplines into four broad categories, including Natural Sciences, Applied Sciences, Social Sciences, and Humanities, following the OECD Fields of Research and Development (FORD) classification scheme~\cite{manual2015frascati}. 

Specifically, Natural Sciences encompasses \textit{Biology, Chemistry, Environmental Science, Geology, Mathematics, Medicine, and Physics}; Applied Sciences includes \textit{Computer Science, Engineering, and Materials Science}; Social Sciences covers \textit{Business, Economics, Geography, Political Science, Psychology, and Sociology}; while Humanities consists of \textit{Art, History, and Philosophy}.

To perform robustness and stringency tests, we supplement these data with the AMiner dataset consisting of $690,972$ Computer Science papers~\cite{tang2008arnetminer}, which provides (i) high-quality disambiguated author information needed to accurately reconstruct the historical collaboration network and (ii) the NSF funding statements needed to identify relevant papers for the policy analysis.

Data filtering criteria are as follows: only ``Research Articles'' are retained, excluding reviews and editorials; only samples whose team members have traceable publication records within the 5 years prior to the focal paper's publication are retained to ensure the complete construction of historical collaboration networks (Robustness checks for time windows of 3-7 years show consistent results; see details in \textit{SI Appendix}, Table S3).

\subsection*{Statistical Analysis and Variable Measurement}
We employ the high-dimensional fixed-effects Ordinary Least Squares (OLS) regression models to examine the statistical relationship between team structural diversity and disruptive innovation output, while controlling for a range of confounding variables defined below.

\subsubsection*{CD Index of Disruptive Innovation}
This is our key dependent variable, measured using the Consolidation-Disruption (CD) Index (range: -1 to 1). Derived from the algorithm proposed by Funk and Owen-Smith~\cite{funk2017dynamic}, the CD Index is calculated based on a focal paper's citation network structure: if subsequent literature cites only the focal paper and ignores its references, it is judged as disruptive (positive CD Index); if subsequent literature cites both the focal paper and its references, it is judged as consolidating (negative value). The CD Index captures the net balance of these different citation behaviors within a given time window (see \textit{SI Appendix}, Fig. S1 for an illustrative example of its calculation). We use the field-normalized CD Index with a 5-year citation window in our analysis, calculated as the Z-score based on the mean and standard deviation of CD Index for papers published within the same year and academic discipline (defined as 19 top-level fields in the OpenAlex database).

\subsubsection*{Structural Diversity}
Our independent variable operates at the individual paper-level, which is defined as the number of Connected Components (CCs) in the focal team's historical co-authorship network, normalized by team size ($SD = CC / TeamSize$). Network construction is based on the collaboration records of all co-authors between $t-5$ and $t-1$ years, where $t$ is the publication year of the focal paper (results are consistent using different window sizes from 2 to 7; see details in \textit{SI Appendix}, Table S3). Our SD metric falls in the range $(0, 1]$, with a larger value indicating a higher degree of historical separation among team members before collaborating on the focal paper.

\subsubsection*{Control Variables}
We control for confounding variables related to linguistic content, author experience, and team characteristics. (i) Team attributes include team size (the total number of authors) and team freshness, defined as the proportion of team members who have not collaborated with any other member before (i.e., node freshness~\cite{zeng2021fresh}). (ii) Author-level factors are based on the last author of each paper, who is often considered and used as the senior and experienced member of the team~\cite{sekara2018chaperone, peng2021acceptance}. We include three variables: the number of publications (counted based on the total number of papers up to the publication year of the focal paper; log-transformed), institution's h-index (based on the last author's first-listed affiliation), and author career age (years since first publication). We also include a squared term for author career age to capture the non-linear relationship between author career stage and innovation behaviors in the academic life cycle~\cite{cui2025agingnarrowingscientificinnovation, peng2025gender}. 

(iii) Content controls include the total number of words in title, the readability score of title (measured with Flesch Reading Ease~\cite{flesch1948new}), and the percentage of promotional words in title (adjectives such as ``unique'', ``crucial'', and ``unprecedented''; see the full lexicon of 139 scientific promotional words in~\cite{peng2024promotional}). 
(iv) Besides author, team, and content controls, we include fixed effects for year and 19 major disciplines defined in the OpenAlex database.

\subsection*{Robustness Tests based on Causal Inference Frameworks}
To address covariate unbalance and endogeneity concerns and the possibility of unobserved confounding factors potentially biasing our results, we adopt two causal inference strategies to test the robustness of the observed predictive power of SD on disruptive innovation.

\subsubsection*{Propensity Score Matching (PSM)} Logistic regression is used to estimate the propensity score of a team having high SD score using all control variables used in our main analysis. The one-to-one nearest neighbor matching is used to construct counterfactual samples to compare the difference in CD Index between the treatment group (high SD) and the control group (low SD).
\subsubsection*{Quasi-Natural Experiment} We utilize the NSF policy implemented in 2012 which expanded the eligibility for multi-institutional project funding. We consider this policy-induced change as an exogenous shock to the existing research ecosystem which may impact both the SD and innovation output. No other significant external factors other than this NSF policy were observed during our studied period. Based on NSF-funded papers (N = $41,458$) in the AMiner dataset, a pre-post design is adopted to test the positive association between SD and disruptive innovation (field-normalized CD Index) before and after the policy implementation (2010-2011 vs 2012-2013).

\subsection*{Mediation Analysis}
We test Disciplinary Integration (DI) as one potential mechanism mediating the association between SD and CD Index. We quantify DI of each paper based on the Simpson Diversity Index ($1 - \Sigma p_i^2$), where $p_i$ denotes the proportion of references from discipline $i$ cited in the paper. This metric captures a focal paper's distributional heterogeneity of knowledge inputs rather than a simple counting of different disciplines. We employ a standard Causal Mediation Analysis framework implemented in the Python package statsmodels (version 0.14.0) to estimate the indirect effect of SD on disruptive innovation via the DI path.

\subsection*{Data Availability}
All data and materials needed to reproduce main findings reported in our paper are documented in the paper and the Supplementary Material. The OpenAlex bibliographic database used in this study can be publicly accessed at https://openalex.org/. The AMiner dataset can be accessed at https://www.aminer.cn/. Due to the massive scale of the processed OpenAlex dataset and the excessive size of the derived files (exceeding 1.2 TB, with a computation time of over 2 months), it is not feasible to deposit these data in a standard public repository for direct reproduction. However, our processed AMiner dataset is deposited on Figshare at https://doi.org/10.6084/m9.figshare.31288000. Our code can be accessed on Github at https://github.com/Yichun-Peng/Structural-Diversity-Innovation.

\section*{Declarations}

\subsection*{Acknowledgements}
This work was supported by the National Natural Science Foundation of China (Grant No. 72293575) and the City University of Hong Kong Startup Grant (No. 9610703 to H.P.).

\subsection*{Author Contributions}
Y.P. developed the code, performed data analysis and visualization, and wrote the original manuscript; S.H. and P.Z. conceived the original idea, assisted in study design, data interpretation, and manuscript revision; K.Z., Y.Y., N.Z., and Q.Z. provided critical feedback and contributed to the editing of the manuscript; D.D.Z. designed the study; H.P. supervised the project, refined the theoretical framework, and revised the manuscript.

\subsection*{Competing Interests}
The authors declare no competing interest.

\bibliographystyle{unsrt} 
\bibliography{pnas-sample}

\end{document}